\documentclass[aps,prc,preprint,showpacs,superscriptaddress,amsmath,amssymb,floatfix,nofootinbib]{revtex4}
\usepackage[dvips]{graphicx,color}% Include figure files
\usepackage{dcolumn}% Align table columns on decimal point
\usepackage{bm}% bold math

\def\pmb#1{\setbox0=\hbox{#1}%
  \kern-.025em\copy0\kern-\wd0 
  \kern.05em\copy0\kern-\wd0
  \kern-.025em\raise.0433em\box0 }

\def\lambdabar{\protect\@lambdabar}
\def\@lambdabar{%
\relax
\bgroup
\def\@tempa{\hbox{\raise.73\ht0
\hbox to0pt{\kern.25\wd0\vrule width.5\wd0
height.1pt depth.1pt\hss}\box0}}%
\mathchoice{\setbox0\hbox{$\displaystyle\lambda$}\@tempa}%
{\setbox0\hbox{$\textstyle\lambda$}\@tempa}%
{\setbox0\hbox{$\scriptstyle\lambda$}\@tempa}%
{\setbox0\hbox{$\scriptscriptstyle\lambda$}\@tempa}%
\egroup
}
\begin{document}

\preprint{J-PARC-TH-230}

\title{\boldmath
$\Xi$-nucleus potential for $\Xi^-$ quasifree production in the $^9$Be($K^-$,~$K^+$) reaction
%$\Xi^-$ quasifree production spectra in the $^9$Be($K^-$,~$K^+$) reaction
}% Force line breaks with \\

\author{Toru~Harada}% 
\email{harada@osakac.ac.jp}
\affiliation{%
Research Center for Physics and Mathematics,
Osaka Electro-Communication University, Neyagawa, Osaka, 572-8530, Japan
}%Lines break automatically or can be forced with \\
\affiliation{%
J-PARC Branch, KEK Theory Center, Institute of Particle and Nuclear Studies,
High Energy Accelerator Research Organization (KEK),
203-1, Shirakata, Tokai, Ibaraki, 319-1106, Japan
}%Lines break automatically or can be forced with \\
\author{Yoshiharu~Hirabayashi}%
%\email{hirabay@iic.hokudai.ac.jp}
\affiliation{%
Information Initiative Center, 
Hokkaido University, Sapporo, 060-0811, Japan
}%Lines break automatically or can be forced with \\

\date{\today}% It is always \today, today,

\begin{abstract}
We study phenomenologically a $\Xi^-$ production spectrum of the $^9$Be($K^-$,~$K^+$) 
reaction at 1.8 GeV/$c$ within the distorted-wave impulse approximation 
using the optimal Fermi-averaged $K^-  p \to K^+ \Xi^-$ amplitude. 
We attempt to clarify properties of a $\Xi$-nucleus potential for $\Xi^-$-$^8$Li, 
comparing the calculated spectrum with the data of the BNL-E906 experiment. 
The results show a weak attraction in the $\Xi$-nucleus potential for $\Xi^-$-$^8$Li, 
which can sufficiently explain the data in the $\Xi^-$ quasifree region.
The strength of $V_0^\Xi =$ $-17 \pm 6$ MeV is favored within the Woods-Saxon potential,
accompanied by the reasonable absorption of $W_0^\Xi =$ $-5$ MeV for
 $\Xi^-p \to \Xi^0n$, $\Lambda\Lambda$ transitions in nuclear medium.
It is difficult to determine the value of $W_0^\Xi$ from the data 
due to the insufficient resolution of 14.7 MeV FWHM.
The energy dependence of the Fermi-averaged $K^-  p \to K^+ \Xi^-$ 
amplitude is also confirmed by this analysis, 
and its importance in the nuclear ($K^-$,~$K^+$) reaction is emphasized.

\end{abstract}
\pacs{21.80.+a, 24.10.Eq, 25.80.Hp, 27.20.+n 
}

\keywords{Hypernuclei, Resonant state, Sigma-nucleus potential
}%Use showkeys class option if keyword
\maketitle

%-------------------------------------------------
% Text 

\section{Introduction}
\label{Intro}

Recently, Nakazawa {\it et al.}~\cite{Nakazawa15} reported the first evidence of a bound state 
of the $\Xi^-$-$^{14}$N system which was identified by the ``KISO'' event in the KEK-E373 experiment.
This result supports that the $\Xi$-nucleus potential has a weak attraction of 
$V_\Xi\simeq$ $-14$ MeV
in the Wood-Saxon (WS) potential, as suggested by previous analyses 
\cite{Tadokoro95,Fukuda98,Khaustov00}.
However, there still remains an uncertainty about the nature of the $S=-2$ dynamics 
caused by $\Xi N$ interaction and $\Xi N$-$\Lambda\Lambda$ coupling in nuclei 
due to the limit to the available data. 
More experimental information is needed for the understanding of $\Xi$ hypernuclei.  
Recently, Nagae {\it et al.}~\cite{Nagae18} have performed an accurate 
observation of the $\Xi^-$ production spectrum in double-charge exchange reactions 
($K^-$,$K^+$) on $^{12}$C targets at 1.8 GeV/$c$ in the J-PARC E05 experiment, 
and their analysis is now ongoing.
The double-charge exchange reactions such as ($K^-$,~$K^+$) on nuclear targets provide
to produce neutron-rich $\Xi$ hypernuclei, e.~g., the neutron excess of 
$(N-Z)/(N+Z)=$ 0.25 for a $\Xi^-$-$^8$Li system, which is populated on $^9$Be.
The behavior of the $\Xi^-$ in the neutron-excess environment is strongly connected with 
the nature of neutron stars \cite{Chatterjee16} 
in which the baryon fraction is found to depend on properties of hypernuclear 
potentials \cite{Balberg97}.

Kohno \cite{Kohno19} examined theoretically 
$\Xi^-$ production spectra for the quasifree (QF) interaction region 
in the ($K^-$,$K^+$) reactions on $^9$Be and $^{12}$C targets 
in the semiclassical distorted wave method, using the $\Xi$-nucleus 
potential derived from the next-to-leading order (NLO) 
in chiral effective field theory.
However, it has shown that the calculated $\Xi^-$ QF spectrum on $^9$Be 
seems to be insufficient to reproduce the experimental data, 
so that quantitative information on the $\Xi$-nucleus 
potential for $\Xi^-$-$^8$Li ($^9_{\Xi^-}$He) may be inreliable.

In this paper, we investigate phenomenologically the $\Xi^-$ QF spectrum 
produced via the $^9$Be($K^-$,~$K^+$) reaction at 1.8 GeV/$c$ 
in order to extract valuable information on the $\Xi$-nucleus (optical) potential 
for the $\Xi^-$-$^8$Li system from the data of 
the BNL-E906 experiment \cite{Kohno19,Tamagawa00}.
We attempt to clarify properties of the $\Xi$-nucleus potential for $\Xi^-$-$^8$Li 
and to understand a mechanism of the $\Xi^-$ QF spectrum 
in comparison with the data \cite{Tamagawa00}.
Thus we demonstrate the calculated $\Xi^-$ QF spectrum in the $^9$Be($K^-$,~$K^+$) reaction 
within the distorted-wave impulse approximation (DWIA), taking into account 
the energy dependence of the $K^-p \to K^+ \Xi^-$ amplitude in the optimal 
Fermi-averaging procedure \cite{Harada20,Harada04}.

\section{Calculations}
\label{Cal}

\subsection{Distorted-wave impulse approximation}
\label{DWIA}

Let us consider production of $\Xi$ hypernuclear states in 
the nuclear ($K^-$,~$K^+$) reaction.
According to the Green's function method \cite{Morimatsu94} in the DWIA, 
an inclusive $K^+$ double-differential laboratory cross section 
of the $\Xi^-$ production on a nuclear target with a spin 
$J_A$ (its $z$-component $M_A$) \cite{Hufner74,Auerbach83,Dover80}
is given by
\begin{equation}
{{d^2\sigma} \over {d\Omega dE} } 
 = {1 \over {[J_A]}} \sum_{M_A}S(E) 
\label{eqn:e3}
\end{equation}
with $[J_A]=2J_A+1$. The strength function $S(E)$ is 
written as
\begin{eqnarray}
S(E)&=&-{1 \over \pi}{\rm Im} \sum_{\alpha \alpha'}
\int d{\bm r}d{\bm r'}
F_{\Xi}^{\alpha \, \dagger}({\bm r}) 
{G}_{\Xi}^{\alpha\alpha'}(E;{\bm r},{\bm r}')\nonumber\\
&& \times 
F_{\Xi}^{\alpha'}({\bm r}'),
\label{eqn:e4}
\end{eqnarray}
where 
${G}_{\Xi}^{\alpha\alpha'}$ is a complete Green's function for 
a $\Xi$ hypernuclear system, 
$F_{\Xi}^{\alpha}$ is a $\Xi$ production amplitude
defined by
\begin{equation}
  F_{\Xi}^{\alpha} = 
  \beta^{1 \over 2}{\overline{f}}_{K^-p \to K^+\Xi^-}
  \chi_{{\bm p}_{K^+}}^{(-) \ast}
  \chi_{{\bm p}_{K^-}}^{(+)} 
  \langle \alpha \, | \hat{\mit\psi}_p | \, {\mit\Psi}_A \rangle,
\label{eqn:e5}
\end{equation}
and $\alpha$ ($\alpha'$) denotes the complete set of eigenstates for the system.
The kinematical factor $\beta$ denotes the translation 
from a two-body $K^-$-$p$ laboratory system to a $K^-$-nucleus 
laboratory system. 
$\overline{f}_{K^-p \to K^+\Xi^-}$ is a Fermi-averaged amplitude 
for the $K^-p \to K^+\Xi^-$ reaction in nuclear medium \cite{Auerbach83,Dover80,Harada04}.
$\langle \alpha \, | \hat{\mit\psi}_p  | \, {\mit\Psi}_A \rangle$ 
is a hole-state wave function for a struck proton in the target.
$\chi_{{\bm p}_{K^+}}^{(-)}$ and $\chi_{{\bm p}_{K^-}}^{(+)}$ 
are distorted waves for outgoing $K^+$ and 
incoming $K^-$ mesons, respectively.
The laboratory energy and momentum transfers are 
$\omega= E_{K^-}-E_{K^+}$ and ${\bm q}={\bm p}_{K^-} - {\bm p}_{K^+}$, respectively; 
$E_{K^+}$ and ${\bm p}_{K^+}$ ($E_{K^-}$ and ${\bm p}_{K^-}$) denote 
an energy and a momentum of the outgoing $K^+$ (incoming $K^-$),
respectively. 

Due to a high momentum transfer 
$q \simeq$ 390--600 MeV/$c$ in the nuclear ($K^-$,~$K^+$) reaction 
for $K^+$ forward-direction angles of $\theta_{\rm lab}=$ 1.5$^\circ$--8.5$^\circ$ 
at $p_{K^-}=$ 1.8 GeV/$c$, 
we simplify the computational procedure for $\chi_{{\bm p}_{K^+}}^{(-)}$ 
and $\chi_{{\bm p}_{K^-}}^{(+)}$, 
using the eikonal approximation \cite{Dover80}. 
To reduce ambiguities in the distorted-waves, 
we adopt the same parameters used in calculations for the
$\Lambda$ and $\Sigma^-$ QF spectra in nuclear ($\pi^\pm$,~$K^+$) 
and ($K^-$, $\pi^\pm$) reactions \cite{Harada04,Harada05,Harada18}.
Here we used the total cross sections of $\sigma_{K^-}$= 28.9 mb 
for the $K^- N$ scattering and $\sigma_{K^+}$= 19.4 mb for the $K^+ N$ scattering, 
and $\alpha_{K^-} = \alpha_{K^+} =$ 0, as the distortion parameters.
We also took into account the recoil effects, which are very important to estimate 
the hypernuclear production cross section for a light nuclear system \cite{Harada19}, 
leading to an effective momentum transfer having 
$q_{\rm eff} \simeq (1-1/A)q \simeq 0.80 q$ for $A=$ 9.

Recently, the authors \cite{Harada20} have found the strong energy dependence of 
the $K^-p \to K^+\Xi^-$ reaction in the nuclear medium,  
together with the angular dependence for $\theta_{\rm lab}$. 
Therefore, we emphasize that 
such behavior of $\overline{f}_{K^-p \to K^+\Xi^-}$ 
plays a significant role in explaining the shape 
of the spectrum in the nuclear ($K^-$,~$K^+$) reaction \cite{Harada20}
as well as those in the nuclear ($\pi^\pm$,~$K^+$) reactions \cite{Harada04,Harada05,Harada18}.
Because $\overline{f}_{K^-p \to K^+\Xi^-}$ provides to modify the spectral 
shape including the $\Xi^-$ QF region widely, 
thus one must extract carefully information concerning 
the $\Xi$-nucleus potential from the data.

\subsection{Wave functions}
\label{Wave}

For the $^{9}$Be target, 
the single-particle (s.~p.) description of protons is assumed for simplicity.
We simulate the calculated results of the s.~p.~energies of the nucleons and 
the root-mean-square (rms) radius of $\langle r_N^2 \rangle^{1/2}$ 
for their wave functions 
in the liner combination of atomic orbits (LCAO) models \cite{Okabe77}
which well describe the ground state of $^{9}$Be($3/2^-_{\rm g.s.}$; $T=$ 1/2) 
as $\alpha+\alpha+n$ clusters.
Thus we compute the s.~p.~wave functions for the protons in $0p$ and $0s$, 
using the WS potential with $R=r_0A^{1/3}$ and $a=$ 0.67 fm and 
omitting the spin-orbit potential; the strength parameter of the potential 
is adjusted to be $V^N_0=$ $-$58.0 MeV, 
together with the size parameter of $R=1.60A^{1/3}=$ 3.33 fm 
which may be rather large due to the structure of $\alpha+\alpha+n$. 
Here we obtain the s.~p.~energies of $-$22 MeV for $0p_{1/2}$ and $-$35 MeV for $0s_{1/2}$, 
which are consistent with the data of the proton separation energies in $^{9}$Be($p$,~2$p$) 
reactions indicating widths of 8 MeV for $0p$ and 13 MeV for $0s$ \cite{Tilley02,Jacob66}. 
The charge radius of $^9$Be(3/2$^-_{\rm g.s.}$) is estimated to be 2.53 fm, 
which is in good agreement with the data of $2.519 \pm 0.012$ fm in electron elastic 
scatterings on $^9$Be \cite{Vries87}. 
Note that we must tune in the energies of the s.~p.~states for the protons
as well as the matter rms radius of $\langle r_N^2 \rangle^{1/2}$ for their wave functions, 
leading to the fact that the shape of the calculated QF spectrum  
in the ($K^-$,~$K^+$) reaction sufficiently explain the data.

To calculate the $\Xi^-$ QF spectrum in the nuclear ($K^-$,~$K^+$) spectrum 
within the DWIA, we use the Green's function method \cite{Morimatsu94}, 
which is one of the most powerful treatments in the calculation 
of a spectrum describing not only bound states but also 
continuum states with an absorptive potential 
for spreading components.
Because non-spin-flip processes seem to dominate 
in the $K^- p \to K^+ \Xi^-$ reaction at 1.8 GeV/$c$ \cite{Sharov11}, 
hypernuclear configurations of $[J_C^\pi \otimes j_\Xi^\pi]_{J_B^\pi}$ 
with 
$J_B^\pi=$ 3/2$^+$, 5/2$^+$, 1/2$^-$, 3/2$^-$, 5/2$^-$, 7/2$^-$, 3/2$^-$, 5/2$^-$, $\cdots$, 
are populated in $^9_{\Xi^-}$He with $T_B=$3/2;
we take the $^8$Li core nucleus states with $J_C^\pi=$ 2$^+$, $1^+$, $2^-$, and $1^-$ 
that are given in 
($3/2^- \otimes p^{-1}_{3/2,1/2}$)$_{2^+,1^+}$ and 
($3/2^- \otimes s^{-1}_{1/2}$)$_{2^-,1^-}$ configurations 
formed by a proton-hole state on $^{9}$Be($3/2^-_{\rm g.s.}$), 
and the $\Xi^-$ with $j_\Xi^\pi=\ell_\Xi\otimes 1/2=$ 1/2$^+$, 3/2$^-$, 1/2$^-$, $\cdots$ 
that are given in $\ell_\Xi \leq 15$ being enough to converge 
in calculations for the $\Xi^-$ spectrum. 
Here the components of $\Xi^0n$ and $\Lambda\Lambda$ channels are not considered explicitly  
because the $\Xi^-p \to \Xi^0n$, $\Lambda\Lambda$ transition processes 
may be described as a spreading imaginary potential in $\Xi$ bound and continuum regions.

% Figure 1
%%%%%%%%%%%%%%%%%%%%%%%%%%%%%%%%%%%%%%%%%%%%%%%%%%%%%%%%%%%%%%%%%
\begin{figure}[tb]
  % fig1
  \begin{center}
  \includegraphics[width=1.\linewidth]{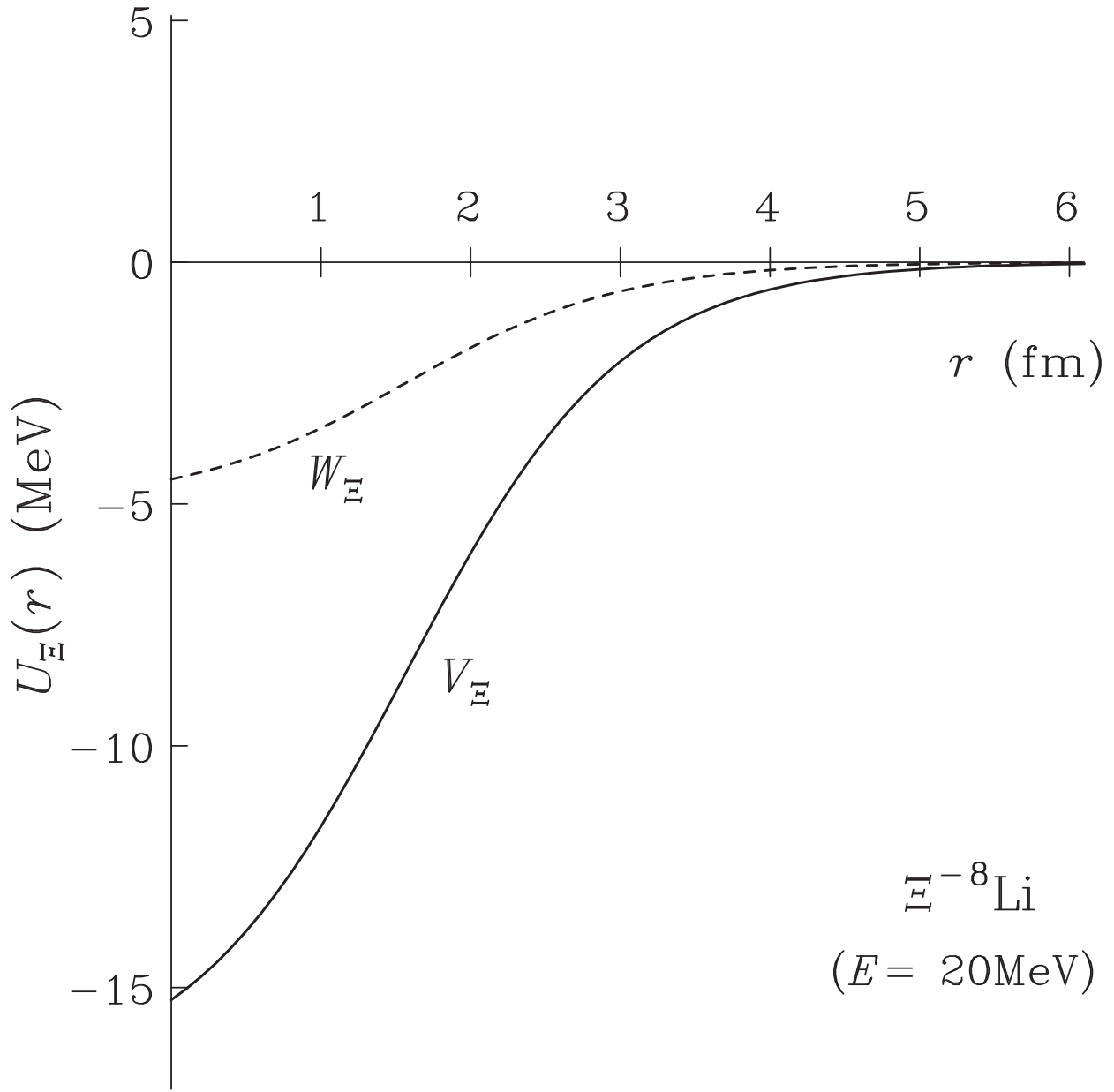}
  \end{center}
  \caption{\label{fig:1}
  Real and imaginary parts of the $\Xi$-nucleus potential 
  $U_{\Xi}$ for $\Xi^-$-$^{8}$Li at the energy $E=$ 20 MeV, as a function of the distance 
  between the $\Xi^-$ and the $^8$Li nucleus. 
  Solid and dashed curves denote the calculated values for $V^{\Xi}_0=$ $-17$ MeV and 
  for $W^{\Xi}_0=$ $-5$ MeV in the WS form, respectively, 
  using $R=$ $r_0A_{\rm core}^{1/3}=$ 1.57 fm where $r_0=$ 0.783 fm 
  and $a=$ 0.722 fm. 
  }
\end{figure}
%%%%%%%%%%%%%%%%%%%%%%%%%%%%%%%%%%%%%%%%%%%%%%%%%%%%%%%%%%%%%%%%%

\section{$\Xi$-nucleus potential}
\label{pot}

The $\Xi$-nucleus final states are obtained by solving 
the Schr\"odinger equation
\begin{eqnarray}
&& \left[-{\hbar^2 \over 2\mu}\nabla^2 
+ U_\Xi(r)+ U_{\rm Coul}(r)  \right] \Psi_\Xi 
= E \Psi_\Xi,
\label{eqn:e4}
\end{eqnarray}
where $\mu$ is the $\Xi$-nucleus reduced mass, 
$U_\Xi$ is the $\Xi$-nucleus potential, and $U_{\rm Coul}$ is the Coulomb potential.
The $\Xi$-nucleus potential for $\Xi^-$-$^8{\rm Li}$ is given by
\begin{eqnarray}
U_{\Xi}(r)&=& V_{\Xi}(r)+iW_{\Xi}(E,r)  \nonumber\\
          &=&[V_0^{\Xi}+iW_0^{\Xi}g(E)]f(r)
\label{eqn:e5}
\end{eqnarray}
with the assumption of the WS form
\begin{eqnarray}
f(r)=[1 + \exp{\{(r-R)/a \}}]^{-1},
\label{eqn:e6}
\end{eqnarray}
where $R= r_0A^{1/3}_{\rm core}$ and $a$ denote a radius 
and a diffuseness of the potential, respectively. 
$V_0^{\Xi}$ is a strength parameter for the real part of the potential;
$W_0^{\Xi}$ is a strength parameter for the imaginary part of the potential, 
which denotes the $\Xi^-$ absorption processes including 
the $\Xi^-p \to \Xi^0n$, $\Lambda\Lambda$ reactions.
$g(E)$ is an energy-dependent function which increases linearly 
from 0.0 at $E=E_{\rm th}(\Lambda)$ to 1.0 at $E=$ 20 MeV
with respect to $\Xi^-$ threshold, 
as often used in nuclear optical models \cite{Yamamoto88},
where $E_{\rm th}(\Lambda)=$ $-23.3$ MeV corresponds to 
the $\Lambda$ emitted threshold.

The ground state of $^8$Li(2$^+_{\rm g.s.}$) has a bound state 
at the neutron binding energy of $B_n=$ 2.03 MeV 
with respect to the $n + {^7{\rm Li_{g.s.}}}$ threshold \cite{Tilley02}; 
the matter rms radius of $\langle r^2_{\rm m} \rangle^{1/2}=$ 2.39 $\pm$ 0.06 fm
is observed experimentally.  
Thus the appropriate parameters of ($r_0$,~$a$) in Eq.~(\ref{eqn:e6}) must be used, 
as we shall mention below. 

To determine the parameters of ($r_0$,~$a$) for the nuclear core 
in the WS form, we adopt a folding-model potential obtained by 
convoluting the nuclear one-body density for $^8$Li with a two-body $\Xi^- N$ force.
We assume the s.p.~density of the spherical shell model for simplicity; 
the modified harmonic oscillator (MHO) model is used in a systematic 
description of a size and a density distribution for Li isotopes 
with $A=$ 6--9 \cite{Tanihata13}.
For $^8$Li(2$^+_{\rm g.s.}$), we choose carefully the MHO size parameters 
of $b_s=$ 1.42 fm and $b_p=$ 1.95 fm 
with center-of-mass and nucleon-size corrections, adjusting the matter rms 
radius of $\langle r^2_{\rm m} \rangle^{1/2}=$ 2.39 fm \cite{Tanihata13}.
Following to the procedure in Ref.~\cite{Harada18}, 
we use the WS form with the parameters of ($r_0$,~$a$) adjusted 
to give a best least-squares fit to the radial shape of the form factor 
obtained by folding a gaussian range of $a_{\Xi N}=$ 1.2 fm 
into the matter MHO density distribution \cite{Harada18}. 
The parameters of the resultant WS form in Eq.~(\ref{eqn:e6}) are 
$r_0=$ 0.783 fm, $a=$ 0.722 fm, and $R=r_0A_{\rm core}^{1/3}=$ 1.57 fm, 
which reproduce the radial shape of the form factor very well;  
the rms radius of the potential denotes
\begin{eqnarray}
\langle r^2 \rangle^{1/2}_V
=\left[\int r^2 V_{\Xi}(r) d{\bm r}\Big/\int V_{\Xi}(r) d{\bm r}\right]^{1/2}= 2.81 \,\mbox{fm}.
\end{eqnarray}

On the other hand, 
the spreading imaginary parts of  $W_0^\Xi$ may represent complicated continuum 
states of ${^{9}_{\Lambda\Lambda}}{\rm He}^*$, $^{9}_{\Xi^-}$He$^*$, 
and $^{9}_{\Xi^0}$He$^*$. 
Considering the states of ${^7{\rm He}(3/2^-_{\rm g.s.})}$ located at $E_{\rm ex}=$ 0.45 MeV 
above the $n+{^6{\rm He}}$ threshold~\cite{Tilley02}, 
we have the $\Lambda$ emitted threshold corresponding 
to the $\Lambda+{^8_\Lambda{\rm He}}$ threshold for the $\Xi^-p \to \Lambda\Lambda$ transition.
The threshold-energy difference between $\Xi^-$-$^{8}$Li and $\Lambda$-$^{8}_\Lambda$He 
channels accounts for 
$\Delta M = M({^8{\rm Li}})+m_{\Xi^-} -M({^8_\Lambda{\rm He}})-m_\Lambda=$ 23.3 MeV, 
where $M({^8{\rm Li}})=$ 7471.4 MeV and $M({^8_\Lambda{\rm He}})=$ 7654.1 MeV.
For the $\Xi^-p \to \Xi^0n$ transition, 
the $\Xi^0$ emitted threshold for $\Xi^0$-$^8{\rm He}$ is located at 
$E=$ 4.3 MeV above the $\Xi^-$-$^{8}$Li threshold.
The spin-orbit potential for $\Xi^-$ is also considered to denote a term of 
$V^{\Xi}_{\rm so}{(1/r)}{[df(r)/dr]}{\bm \sigma}{\cdot}{\bm L}$, 
where $V^{\Xi}_{\rm so}\simeq$ ${1 \over 10}V^{N}_{\rm so}$
$\simeq$ 2 MeV \cite{Bouyssy82}.
For $U_{\rm Coul}$, we use the attractive Coulomb potential 
with the uniform distribution of a charged sphere where
$Z = 3$ for $\Xi^-$-$^8$Li.

We attempt to determine the strength parameters of $V_0^\Xi$ and $W_0^\Xi$ 
in Eq.~(\ref{eqn:e5}) phenomenologically 
in comparison with the data of the $^{9}$Be($K^-$,$K^+$) reaction.
Figure~\ref{fig:1} shows the real and imaginary parts of the $\Xi$-nucleus potential 
for $\Xi^-$-$^8$Li, choosing the reasonable strengths of $V_0^\Xi$= $-$17 MeV 
and $-$5 MeV, as we will discuss it in Sect.~\ref{results}.

\section{Results}
\label{results}

%%%%%%%%%%%%%%%%%%%%%%%%%%%%%%%%%%%%%%%%%%%%%%%%%%%%%%%%%%%%%%%%%%%%%%%%%%%%%%
% Table 1
\begin{table}[b]
\caption{
\label{tab:table1}
The $\chi^2$-fitting for various strength parameters, $V_0^\Xi$ and $W_0^\Xi$, 
in the WS potential with $r_0=$ 0.738 fm and $a=$ 0.722 fm for $\Xi^-$-$^{8}$Li.
The value of $\chi^2/N$ and the renormalization factor $f_s$ are obtained 
by comparing the calculated spectrum with the $N=$ 17 data points 
of the average cross sections of $\bar{\sigma}_{1.5^\circ\mbox{-}8.5^\circ}$ 
for $p_{K^+}=$ 1.07--1.39 GeV/$c$.
The data were taken from Ref.~\cite{Tamagawa00}.
}
\begin{ruledtabular}
\begin{tabular}{rrrc}
\noalign{\smallskip}
     $V_0^\Xi$ & $W_0^{\Xi}$  
     &  \multicolumn{2}{c}{$N=17$ data points} \\ 
\noalign{\smallskip}
                           \cline{3-4}
\noalign{\smallskip} 
     (MeV)        & (MeV)  & $\chi^2/N$  & $f_s$  \\
\noalign{\smallskip}\hline\noalign{\smallskip}
     $+12$   &  \ $0$      &  69.8/17  & 0.988      \\ %6.9840E+01      0.988090E+00
     \ \ $0$ &  \ $0$      &  37.6/17  & 0.964      \\ %3.7571E+01	9.6359E-01
     $-6$    &  \ $0$      &  26.4/17  & 0.951      \\ %2.6366E+01	9.5136E-01
     $-12$   &  \ $0$      &  18.9/17  & 0.939      \\ %1.8940E+01	9.3910E-01
     $-18$   &  \ $0$      &  15.6/17  & 0.927      \\ %1.5645E+01	9.2679E-01
     $-24$   &  \ $0$      &  16.8/17  & 0.914      \\ %1.6833E+01	9.1442E-01
     $-30$   &  \ $0$      &  22.8/17  & 0.902      \\ %2.2838E+01	9.0195E-01
     $+12$   &  \ $-5$     &  58.3/17  & 0.999      \\ %5.8278E+01      0.999363E+00
     \ \ $0$ &  \ $-5$     &  30.7/17  & 0.975      \\ %3.0690E+01	9.7468E-01
     $-6$    &  \ $-5$     &  21.8/17  & 0.962      \\ %2.1764E+01	9.6235E-01
     $-12$   &  \ $-5$     &  16.5/17  & 0.950      \\ %1.6525E+01	9.5000E-01
     $-17$   &  \ $-5$     &  15.2/17  & 0.940      \\ %1.5210E+01	9.3968E-01
     $-18$   &  \ $-5$     &  15.3/17  & 0.938      \\ %1.5298E+01	9.3761E-01
     $-24$   &  \ $-5$     &  18.4/17  & 0.925      \\ %1.8399E+01	9.2515E-01
     $-30$   &  \ $-5$     &  26.1/17  & 0.913      \\ %2.6126E+01	9.1260E-01
     $+12$   &  \ $-10$    &  49.0/17  & 1.010      \\ %4.8953E+01      0.101031E+01
     \ \ $0$ &  \ $-10$    &  25.7/17  & 0.985      \\ %2.5706E+01	9.8547E-01
     $-6$    &  \ $-10$    &  18.9/17  & 0.973      \\ %1.8873E+01	9.7306E-01
     $-12$   &  \ $-10$    &  15.6/17  & 0.961      \\ %1.5637E+01	9.6063E-01
     $-18$   &  \ $-10$    &  16.3/17  & 0.948      \\ %1.6293E+01	9.4816E-01
     $-24$   &  \ $-10$    &  21.1/17  & 0.936      \\ %2.1131E+01	9.3563E-01
     $-30$   &  \ $-10$    &  30.4/17  & 0.923      \\ %3.0416E+01	9.2303E-01
\end{tabular}
\end{ruledtabular}
\end{table}

% Figure 2
%%%%%%%%%%%%%%%%%%%%%%%%%%%%%%%%%%%%%%%%%%%%%%%%%%%%%%%%%%%%%%%%%
\begin{figure}[t!]
%\begin{minipage}{0.49\linewidth}
  \begin{center}
  \includegraphics[width=1.00\linewidth]{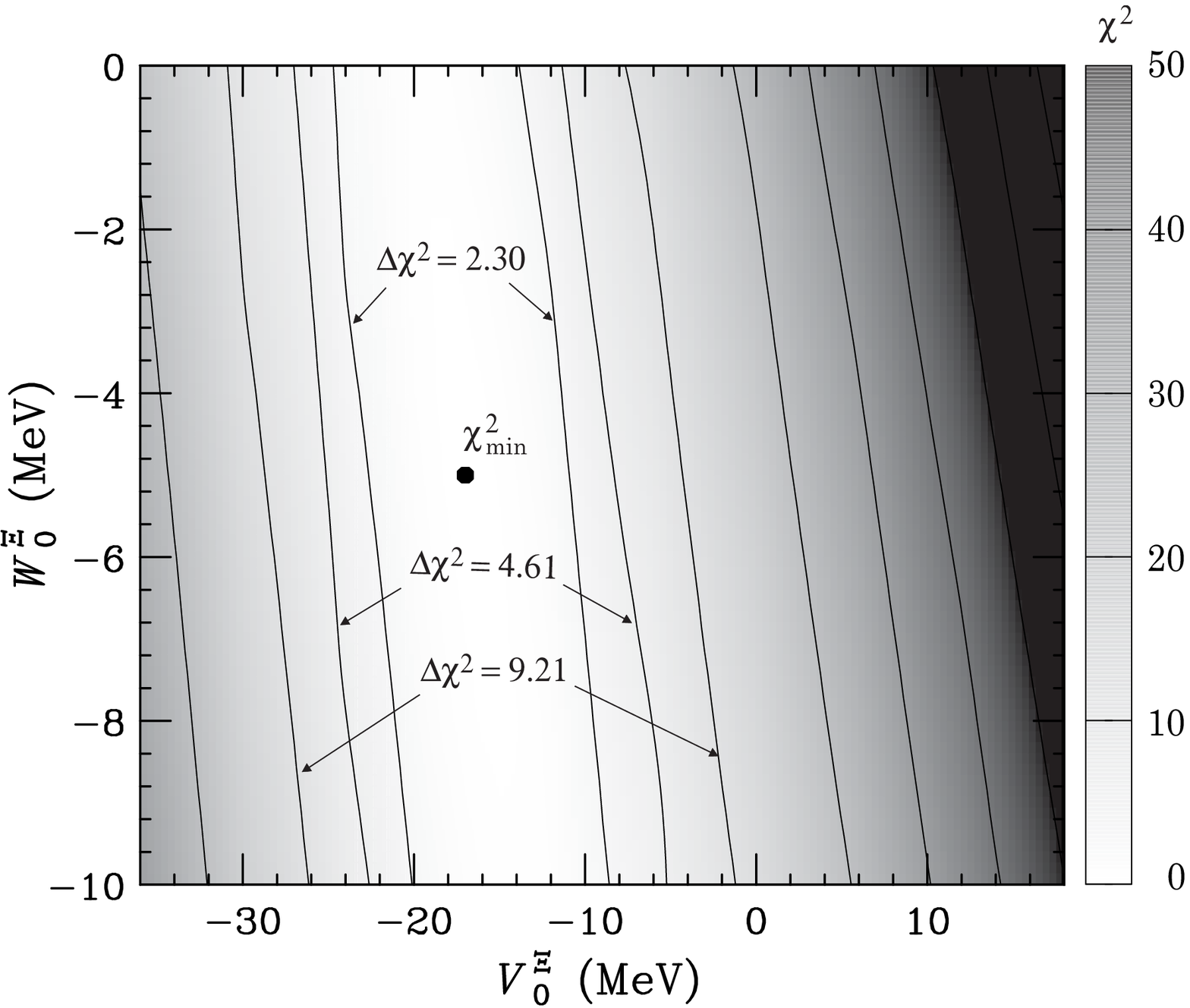}
  \end{center}
%\end{minipage}
  \caption{\label{fig:2}
  Contour plots of the $\chi^2$-value distribution 
  in the \{$V_0^\Xi$, $W_0^\Xi$\} plane from fitting to the average cross section 
  of $\bar{\sigma}_{1.5^\circ\mbox{-}8.5^\circ}$ in the $^9{\rm Be}$($K-$,~$K^+$) 
  reaction at $p_{K^-}=$ 1.8 GeV/$c$. 
  A solid circle denotes the minimum position of $\chi^2_{\rm min}=$ 15.2 at 
 ($V_0^\Xi$,~$W_0^\Xi$) = ($-$17 MeV,$-$5 MeV) with $f_s=$ 0.940. 
  The solid lines labeled by $\Delta \chi^2=$ 2.30, 4.61, and 9.21 
  correspond to 68\%, 90\%, and 99\% confidence levels for 2 parameters, respectively.
}
\end{figure}
%%%%%%%%%%%%%%%%%%%%%%%%%%%%%%%%%%%%%%%%%%%%%%%%%%%%%%%%%%%%%%%%%

\subsection{$\chi^2$ fitting}
\label{fitting}

Tamagawa {\it et al.}~(BNL-E906 collaboration) reported 
the experimental data of  
the $\Xi^-$ QF spectra for the $K^+$ forward-direction angles of 
$\theta_{\rm lab}=$ 1.5$^\circ$--8.5$^\circ$ in the $^9$Be($K^-$,~$K^+$) reactions 
at the incident $K^-$ momentum of $p_{K^-}=$ 1.8 GeV/$c$ \cite{Tamagawa00}.
The average cross section $\bar{\sigma}_{1.5^\circ\mbox{-}8.5^\circ}$ in the laboratory frame
was obtained by 
\begin{equation}
\bar{\sigma}_{1.5^\circ\mbox{-}8.5^\circ} \equiv
\int_{\theta_{\rm lab}= 1.5^\circ}^{\theta_{\rm lab}= 8.5^\circ} \!\!
\left({d^2\sigma \over dp_{K^+}d\Omega_{K^+}} \right) d\Omega
\bigg/
\int_{\theta_{\rm lab}= 1.5^\circ}^{\theta_{\rm lab}= 8.5^\circ} \!\! d\Omega
\label{eqn:e7}
\end{equation}
with the detector resolution of 14.7 MeV FWHM \cite{Tamagawa00}.
The strength parameters of $V_0^\Xi$ and $W_0^{\Xi}$ in Eq.~(\ref{eqn:e5}) should 
be adjusted appropriately to reproduce the data of 
$\bar{\sigma}_{1.5^\circ\mbox{-}8.5^\circ}$.

We consider the $\Xi^-$ QF spectrum for $\Xi^-$-$^8$Li hypernuclear 
states with $J^\pi_B$, $T_B=3/2$, 
using the Green's function method \cite{Morimatsu94},
in order to be compared with the data of the ${^9{\rm Be}}$($K^-$,~$K^+$) reaction
at the BNL-E906 experiment \cite{Tamagawa00}. 
Calculating the spectra for $\theta_{\rm lab}=$ 1.5$^\circ$--8.5$^\circ$, 
we estimate the average cross section 
for the corresponding $\bar{\sigma}_{1.5^\circ\mbox{-}8.5^\circ}$ in Eq.~(\ref{eqn:e7}). 
To make a fit to the spectral shape of the data, 
we will introduce a renormalization factor of $f_s$ 
into the absolute value of the calculated spectrum 
because the eikonal distortion and the amplitude 
of $\overline{f}_{K^-p \to K^+\Xi^-}$ would have some ambiguities~\cite{Dover80,Harada20}. 
The detector resolution of 14.7 MeV FWHM is also taken into account. 
We obtain the values of $\chi^2$ for fits to the data points 
of $N=$ 17 in $p_{K^+}=$ 1.07--1.39 GeV/$c$, 
varying the strengths of ($V_0^\Xi$,~$W_0^\Xi$) and $f_s$;
we assumed the value of 0.018 $\mu$b/sr/MeV$c^{-1}$ as a constant background.
Thus we estimate the average cross section in Eq.~(\ref{eqn:e7}), 
calculating the spectra for $\theta_{\rm lab}=$ 1.5$^\circ$--8.5$^\circ$
in the parameter region of $V_0^\Xi=$ ($-$36)--(+18) MeV by a 6 MeV energy step
and $W_0^\Xi=$ ($-$10)--0 MeV by a 2 MeV energy step. 
The 1 MeV energy step is taken in the estimation near the $\chi^2_{\rm min}$ point.

% Figure 3
%%%%%%%%%%%%%%%%%%%%%%%%%%%%%%%%%%%%%%%%%%%%%%%%%%%%%%%%%%%%%%%%%
\begin{figure}[t]
\begin{minipage}{1.0\linewidth}
\begin{center}
  \includegraphics[width=1.0\linewidth]{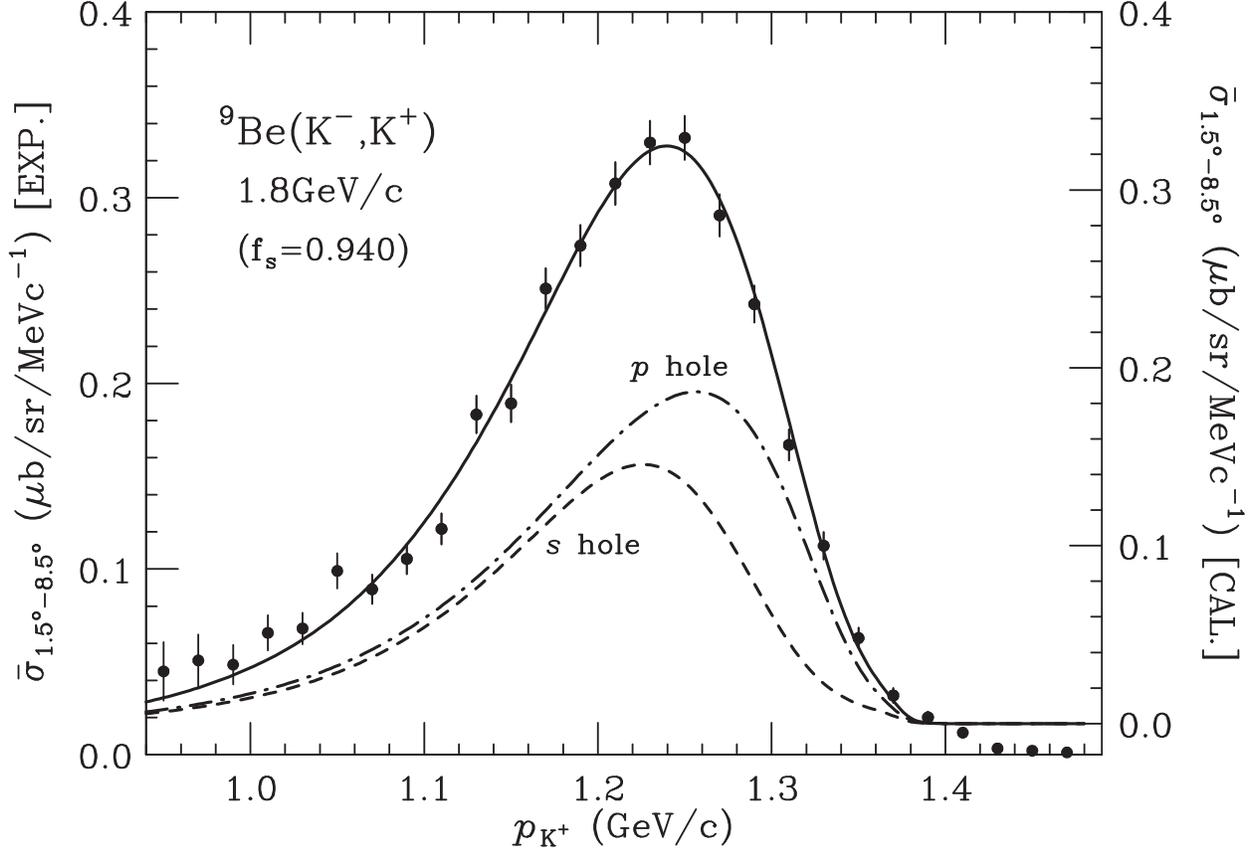}
\end{center}
\end{minipage}
\caption{\label{fig:3}
  Calculated spectrum for $\bar{\sigma}_{1.5^\circ\mbox{-}8.5^\circ}$  
  in the WS potential with 
  $V_0^\Xi =$ $-17$ MeV, $W_0^{\Xi}=$ $-5$ MeV, 
  $r_0=$ 0.738 fm, and $a=$ 0.722 fm, 
  together with the data of the $^{9}$Be($K^-$,~$K^+$) reaction 
  at $p_{\pi^-}$= 1.8 GeV/$c$~\cite{Tamagawa00}. 
  The calculated spectrum is normalized by $f_s=$ 0.940 for fits to the data. 
  Solid, dashed, and dot-dashed curves denote the contributions of total, $s$-hole, 
  and $p$-hole configurations, respectively. 
  The calculated values are folded with a detector resolution of 14.7 MeV FWHM. 
}
\end{figure}
%%%%%%%%%%%%%%%%%%%%%%%%%%%%%%%%%%%%%%%%%%%%%%%%%%%%%%%%%%%%%%%%%%

Figure~\ref{fig:2} displays the contour plots of 
$\chi^2$-value distribution for $\bar{\sigma}_{1.5^\circ\mbox{-}8.5^\circ}$.
The minimum value of $\chi^2$ is found to be $\chi^2_{\rm min}=$ 15.2
at $V_0^\Xi=$ $-$17 MeV, $W_0^\Xi=$ $-5$ MeV, and $f_s=$ 0.940, 
leading to belt-like regions of $\Delta \chi^2 =$ 2.30, 4.61, 
and 9.21 which correspond to 68\%, 90\%, and 99\% confidence levels 
for 2 parameters, respectively, 
where $\Delta \chi^2 \equiv \chi^2 - \chi^2_{\rm min}$.
We find that the value of $\chi^2$ is almost insensitive to $W_0^\Xi$. 
This fact implies that the parameter of $W_0^\Xi$ cannot be determined from 
the BNL-E906 data due to the insufficient resolution of 14.7 MeV FWHM.
Nevertheless, we recognize that the calculated spectrum 
for $V_0^\Xi \simeq$ $-$17 MeV seems to be in good agreement with 
the data when $W_0^\Xi \simeq$ $-$5 MeV; 
it gives the minimum value of $\chi^2/N =$ 15.2/17= 0.89, 
and the standard deviation of $\sigma \simeq$ 6 MeV. 
In Table~\ref{tab:table1}, we list the reduced $\chi^2$ values 
of $\chi^2/N$ in calculations when $V_0^\Xi$= $-$30, $-$24, $-$18, $-$12, $-$6, 0, and $+$12 MeV,  
and $W_0^\Xi$ = $-$10, $-$5, and 0 MeV, comparing the calculated spectra with the data.
Note that the absolute values of the calculated cross section 
can explain the magnitude of the data, as seen by $f_s \simeq$ 0.9--1.0.

Figure~\ref{fig:3} shows the absolute values of the calculated spectrum 
for $\bar{\sigma}_{1.5^\circ\mbox{-}8.5^\circ}$ in the best-fit calculation, 
comparing them with the data of the BNL-E906 experiment at $p_{K^+}=$ 1.07--1.39 GeV/$c$. 
We recognize that an attraction in the $\Xi^-$-$^8$Li potential 
is needed to reproduce the data. 
The contribution of $p$-hole configurations is larger than that of $s$-hole configurations 
in the $\Xi^-$ QF region of $p_{K^+}=$ 1.2--1.4 GeV/$c$, whereas 
the former is similar to the latter in the region of $p_{K^+} <$ 1.2 GeV/$c$ where 
the recoil momentum  grows into $q >$ 540 MeV/$c$. 
Consequently, we confirm that the $\Xi$ potential for $\Xi^-$-$^8$Li has a weak attraction  
in the real part of the WS potential with $r_0=$ 0.738 fm and $a=$ 0.722 fm;
\begin{eqnarray}
&V_0^\Xi = & -17 \pm 6 \ {\rm MeV} \quad \mbox{for $W_0^\Xi=$ $-$5 MeV}.
\label{eqn:e8}
\end{eqnarray}
This potential provides the ability to explain the $^9$Be($K^-$,~$K^+$) data 
at the BNL-E906 experiment.
Several authors \cite{Fukuda98,Khaustov00} attempted to determine 
the values of $V_0^\Xi$ for fits to the shape and magnitude of 
the $\Xi^-$ QF spectra from the data of the $^{12}$C($K^-$,~$K^+$) 
reaction \cite{Khaustov00}.
They suggested that the $\Xi$-nucleus potential has a weak attraction 
of $V_0^\Xi \simeq$ $-14$ MeV in the WS potential. 
It is shown that the results of Eq.~(\ref{eqn:e8}) in our analysis are 
considerably consistent with the results 
of the previous studies \cite{Fukuda98,Khaustov00}.

\subsection{$\Xi^-$-nucleus bound states}

%%%%%%%%%%%%%%%%%%%%
% Table 2
\begin{table*}[bt]
\caption{
\label{tab:table2}
Binding energies $B_{\Xi^-}$ and widths $\varGamma_{\Xi^-}$ 
of the $\Xi^-$-nucleus $(n \ell)$ bound states for 
$\Xi^-$--$^{8}$Li ($^{9}_{\Xi^-}$He).
The strengths of $V_0^\Xi= -17$ MeV and $W_0^\Xi=$ $-2.5$ ($-1.5$) MeV 
are used in the WS potential for the $\Xi^-$ bound region.
These values are estimated in combination with
the $\Xi$-nucleus potential $U_\Xi=V_\Xi+iW_\Xi$
and the Coulomb potential $U_{\rm Coul}$.  
}
\begin{ruledtabular}
\begin{tabular}{cccccccccc}
       & \multicolumn{3}{l}{$V_\Xi+U_{\rm Coul}+iW_\Xi$}
       & \multicolumn{2}{l}{$V_\Xi+U_{\rm Coul}$} 
       & \multicolumn{2}{l}{$V_\Xi$}
       & \multicolumn{2}{l}{$U_{\rm Coul}$} \\ 
\cline{2-4} \cline{5-6} \cline{7-8} \cline{9-10}
$(n \ell)$
   & $-B_{\Xi^-}$ & ${\mit\Gamma}_{\Xi^-}$  & rms  & $-B_{\Xi^-}$ & rms  & $-B_{\Xi^-}$ & rms  & $-B_{\Xi^-}$ & rms  \\
   & (MeV) & (MeV) &  (fm) & (MeV) &  (fm)   & (MeV) &  (fm)  & (MeV) & (fm)   \\
\noalign{\smallskip}\hline\noalign{\smallskip}
\multicolumn{3}{l}{$W_0^\Xi=$ $-2.5$ MeV} & & & & & & &  \\
$1S$ &$-$1.851  &1.118  & 3.95  &$-$1.897  &  3.96 &$-$0.475 & 5.56 &$-$0.255 &  14.6 \\
$2S$ &$-$0.122  &1.3$\times$10$^{-2}$ & 29.3  &$-$0.122  &  29.5 &---      &---   &$-$0.066 &  53.4 \\
$2P$ &$-$0.068  &3.1$\times$10$^{-4}$ & 43.4  &$-$0.068  &  43.4 &---      &---   &$-$0.067 &  44.0 \\ 
\noalign{\smallskip}
\multicolumn{3}{l}{$W_0^\Xi=$ $-1.5$ MeV} & & & & & & &  \\
$1S$ &$-$1.880  &0.669  & 3.95 \\
$2S$ &$-$0.122  &8.0$\times$10$^{-3}$ & 29.4 \\
$2P$ &$-$0.068  &1.9$\times$10$^{-4}$ & 43.4 \\ 
\end{tabular}                                            
\end{ruledtabular}
\end{table*}
%%%%%%%%%%%%%%%%%%%%

In Table~\ref{tab:table2}, we show the numerical results of 
binding energies and widths of the $\Xi^-$-nucleus $(n\ell)$ bound states for $\Xi^-$-$^8{\rm Li}$, 
where $(n\ell)$ denote the principal and angular momentum quantum numbers for the relative 
motion between $\Xi^-$ and $^8{\rm Li}$.
By solving the Schr\"odinger equation of Eq.~(\ref{eqn:e4})
with the WS potential $U_\Xi$ and the finite Coulomb potential $U_{\rm Coul}$, 
we obtain a complex eigenvalue as a Gamow state, 
\begin{eqnarray}
E_{n\ell}= -{B_{\Xi^-}}-i\frac{{\mit\Gamma}_{\Xi^-}}{2}, 
\label{eqn:e8a}
\end{eqnarray}
where $B_{\Xi^-}$ and ${\mit\Gamma}_{\Xi^-}$ denote a binding energy and a width 
of the bound state, respectively. 
When we use $V_0^\Xi=$ $-17$ MeV in the WS potential, we confirm that 
there exists a very shallow $\Xi^-$ $(1S)$ bound state due to the weak attraction 
in the $\Xi$-nucleus potential
even if the Coulomb potential is switched off; 
the binding energy accounts for $B_{\Xi^-}(1S)=$ $0.475$ MeV
and the rms radius of $\langle r^2 \rangle^{1/2}=$ 5.56 fm.
When the Coulomb potential is switched on, 
the binding energy is significantly shifted downward 
in comparison with the corresponding Coulomb eigenstate, as seen in Table~\ref{tab:table2}.
Thus this state is often regarded as a ``Coulomb-assisted'' $\Xi^-$-nucleus bound state; 
$B_{\Xi^-}(1S)=$ $1.90$ MeV and $\langle r^2 \rangle^{1/2}=$ 3.96 fm.  

A $\Xi^-$ hyperon bound in nuclei must be absorbed by strong interaction
via the $\Xi^- p \to \Lambda\Lambda$ conversion process.
To estimate the width of the $\Xi^-$ bound state, 
we assume the value of $W_0^\Xi=$ $-2.5$ MeV, which corresponds to 
the strength of $W_\Xi(E)$ at the $\Xi^-$ threshold ($E=$ 0.0 MeV).
Thus we obtain the width of ${\mit\Gamma}_{\Xi^-}(1S)=$ 1.12 MeV, 
together with $B_{\Xi^-}(1S)=$ $1.85$ MeV. 
When we use $W_0^\Xi=$ $-1.5$ MeV arising from the $\Xi^- p \to \Lambda\Lambda$ conversion
in the $\Xi N$ NLO potential \cite{Haidenbauer16,Haidenbauer19},
we obtain ${\mit\Gamma}_{\Xi^-}(1S)=$ 0.669 MeV 
and $B_{\Xi^-}(1S)=$ $1.88$ MeV. (See also Sect.~\ref{imag}.)

\section{Discussion}
\label{discussion}

% Figure 4
%%%%%%%%%%%%%%%%%%%%%%%%%%%%%%%%%%%%%%%%%%%%%%%%%%%%%%%%%%%%%%%%%
\begin{figure}[t]
\begin{minipage}{1.0\linewidth}
\begin{center}
  \includegraphics[width=1.0\linewidth]{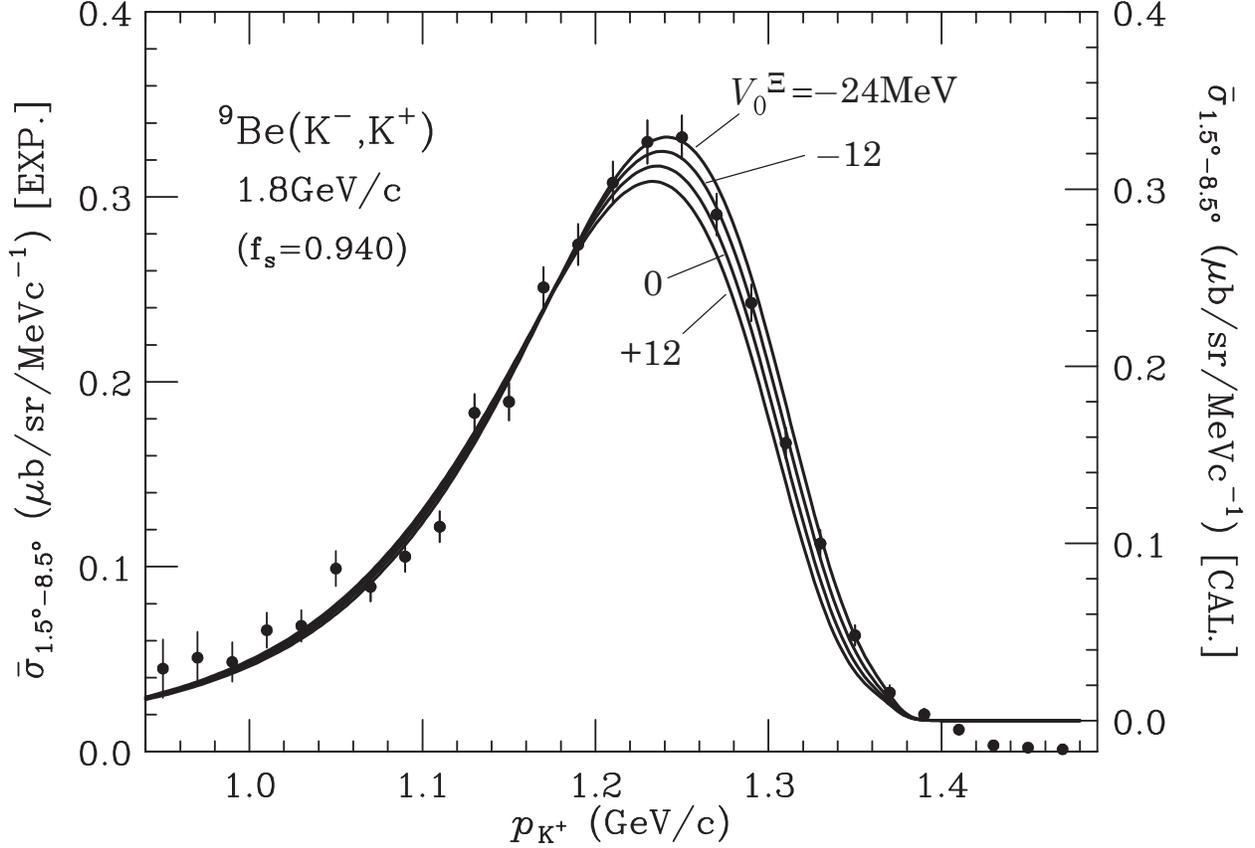}
\end{center}
\end{minipage}
\caption{\label{fig:4}
  Shapes and magnitudes of the calculated spectra for 
  $\bar{\sigma}_{1.5^\circ\mbox{-}8.5^\circ}$ in the $^9$Be($K^-$,~$K^+$) reaction 
  at $p_{K^-}=$ 1.80 GeV/$c$, depending on the strengths of 
  $V_0^\Xi$= $-$24, $-$12, 0, and $+$12 MeV in the WS potential with $W_0^\Xi=$ $-$5 MeV.  
  The spectra are folded with a detector resolution of 14.7 MeV FWHM. 
}
\end{figure}
%%%%%%%%%%%%%%%%%%%%%%%%%%%%%%%%%%%%%%%%%%%%%%%%%%%%%%%%%%%%%%%%%%

\subsection{Effects of the real part of the $\Xi$-nucleus potential}

To see effects of 
the attraction in the $\Xi$-nucleus potential for $\Xi^-$-$^8$Li, 
we discuss the shapes and magnitudes of the calculated spectra.
Figure~\ref{fig:4} shows the absolute values of the calculated 
spectra for $\bar{\sigma}_{1.5^\circ\mbox{-}8.5^\circ}$ 
in the $\Xi^-$ QF region, using various strengths of $V_0^\Xi$. 
We find that the shape and magnitude of the calculated spectrum 
are considerably sensitive to the value of $V_0^\Xi$. 
This confirms that the value of $\chi^2/N$ is significantly 
changed by $V_0^\Xi$. 
The peak position of the QF spectrum is scarcely shifted downward for $p_{K^+}$, 
as $V_0^\Xi$ changes from $-24$ MeV to $+12$ MeV, 
whereas the magnitude of this peak is slightly reduced by 8.4\%.
There appears the difference between the spectra of $V_0^\Xi=$ $(-24)$--$(+12)$ MeV 
in the momentum region of $p_{K^+}>$ 1.2 GeV/$c$, 
corresponding to the region of lower energies $E <$ 140 MeV.
On the other hand, the shapes and magnitudes of the spectra with 
$V_0^\Xi=$ $(-24)$--$(+12)$ MeV become similar to each other 
in the region of higher energies $E >$ 140 MeV ($p_{K^+} < $ 1.2 GeV/$c$).

\subsection{Validity of the imaginary part of the $\Xi$-nucleus potential}
\label{imag}

In Sect.~\ref{fitting}, 
we have found that the shapes and magnitudes of the calculated spectra are not so sensitive 
to the value of $W_0^\Xi$ when we change $W_0^\Xi=$ $(-10)$--$0$ MeV in the imaginary part 
of the $\Xi$-nucleus potential. 
This reason is because a mask of $W_0^\Xi$ is inevitable due to the insufficient 
resolution of 14.7 MeV FWHM. 
Thus we recognize that it is difficult to determine the value of $W_0^\Xi$.

According to the procedure by Gal, Toker, and Alexander \cite{Gal81},
we examine theoretically an appropriate parameter for $W_0^\Xi$ from a viewpoint 
of the first order optical ($t\rho$) potential, 
\begin{eqnarray}
U_\Xi^{(1)}(r)
= t_{\Xi^-p}\rho_{p}(r)+t_{\Xi^-n}\rho_{n}(r)
\label{eqn:e9a}
\end{eqnarray}
in terms of the effective two-body $\Xi N$ elastic $t_{\Xi N}$ scattering matrices 
in the laboratory frame, 
where $\rho_{p,n}(r)$ are the proton and neutron densities of the core nucleus.
By the optical theorem $4\pi{\rm Im}f_{\Xi N}=k_\Xi\sigma_{\rm tot}$ and considering 
collisions of zero energy $\Xi$ with bound nucleons, we obtain
the imaginary part $W^{(1)}_\Xi$ of the optical potential involving 
the $\Xi^-p \to \Xi^0n,\Lambda\Lambda$ conversion, 
which is given by
\begin{eqnarray}
 W^{(1)}_\Xi(r)
&=& -\bigl\langle v_{\Xi^-p}
\sigma(\Xi^-p\to\Xi^0n,\Lambda\Lambda)\bigr\rangle\nonumber\\
&& \times \rho_p(r)/2,
\label{eqn:e9}
\end{eqnarray}
where $v$ is the relative velocity of a $\Xi^-p$ pair,
and $\langle \cdots \rangle$ indicates nuclear medium corrections to the free space 
value of $v\sigma$ arising from Fermi averaging, binding effects, and Pauli principle, etc. 
The cross section is well approximated up to 300 MeV/$c$ in the laboratory system by
the form
\begin{eqnarray}
v_{\Xi^-p}\sigma=(v_{\Xi^-p}\sigma)_0/(1+\alpha v), 
\label{eqn:e10}
\end{eqnarray}
with the two representative parametrization of 
$(v_{\Xi^-p}\sigma)_0=$ 25 mb and $\alpha=$ 18 for 
the $\Xi^-p\to\Xi^0n$,~$\Lambda\Lambda$ reactions, 
fitting to $v_{\Xi^-p}\sigma$ which are given 
by the $\Xi N$ NLO potential \cite{Haidenbauer16,Haidenbauer19}.
Taking into account the closure assumption and nuclear medium corrections \cite{Gal81}, 
we obtain $\langle v_{\Xi^-p}\sigma\rangle=$ 7.02 mb
within the Fermi gas model. 
Using the relation between $\langle v\sigma \rangle$ and ${\rm Im}\,b$, 
where $b$ is the effective parameter of a complex scattering length for $\Xi^-p$, 
we roughly estimate
\begin{eqnarray}
{\rm Im}\,b=\mu\langle v\sigma \rangle/8\pi= 0.078 \, \mbox{fm},
\end{eqnarray}
of which the value corresponds to $W_0^\Xi=$ $-6.2$ MeV in the WS potential. 
We find that this value is similar to $W_0^\Xi=$ $-5$ MeV 
for the minimum value of $\chi_{\rm min}$, as shown in Fig.~\ref{fig:2}.
If we replace the momentum distributions of the Fermi gas model 
by those of the s.~p.~shell model for the finite nuclei, the results may not change. 
Therefore, we believe that the $\Xi$-nucleus potential
with $V_0^\Xi=$ $-17$ MeV and $W_0^\Xi=$ $-5$ MeV is appropriate to 
the study of the $\Xi^-$ QF spectrum in the $^9$Be($K^-$,~$K^+$) reaction 
at $p_{K^-}=$ 1.8 GeV/$c$. 

Considering the same manner 
for only the $\Xi^-p\to\Lambda\Lambda$ conversion \cite{Haidenbauer16,Haidenbauer19}, 
we also obtain $(v_{\Xi^-p}\sigma)_0=$ 4.5 mb and $\alpha=$ 20. 
Thus we estimate ${\rm Im}\,b=$ 0.018 fm, which corresponds to $W_0^\Xi=$ $-1.5$ MeV. 
Such a small absorption of $W_0^\Xi \simeq$ $-1$ MeV 
may be acceptable because the $\Xi^-p \to \Lambda\Lambda$ coupling is recently predicted 
to be rather small \cite{Kohno19,Sasaki20}.

\subsection{Verification of the optimal Fermi-averaged $K^-p\to K^+\Xi^-$ amplitude}

%%%%%%%%%%%%%%%%%%%%
% Table 2
\begin{table*}[t]
\caption{
\label{tab:table3}
Comparison of the standard Fermi-averaged differential cross sections 
$(d\sigma/d\Omega)^{\rm av}_{\rm lab}$
for the $K^-p \to K^+\Xi^-$ reaction at $p_{K^-}=$ 1.8 GeV/$c$
with the differential cross sections 
$(d\sigma/d\Omega)^{\rm free}_{\rm lab}$
for the $K^-p \to K^+\Xi^-$ reaction
in free space \cite{Harada20}. The values are in unit of mb/sr.
}
\begin{ruledtabular}
\begin{tabular}{cccccccccc}
$\theta_{\rm lab}$ 
& $1^\circ$ & $2^\circ$ & $3^\circ$ & $4^\circ$ & $5^\circ$ 
& $6^\circ$ & $7^\circ$ & $8^\circ$ & $9^\circ$ \\
\hline
$(d\sigma/d\Omega)^{\rm av}_{\rm lab}$
& 55.4  & 54.7  & 53.5  &  51.9  &  50.0  & 47.8   & 45.5   &  43.1   & 40.7  \\
$(d\sigma/d\Omega)^{\rm free}_{\rm lab}$
& 67.1  & 65.9  & 64.1  &  61.7  &  58.8  & 55.6   & 52.3   &  49.0   & 45.8  \\
\end{tabular}                                            
\end{ruledtabular}
\end{table*}

%%%%%%%%%%%%%%%%%%%%

% Figure 5
%%%%%%%%%%%%%%%%%%%%%%%%%%%%%%%%%%%%%%%%%%%%%%%%%%%%%%%%%%%%%%%%%
\begin{figure}[t]
\begin{minipage}{1.0\linewidth}
\begin{center}
  \includegraphics[width=1.0\linewidth]{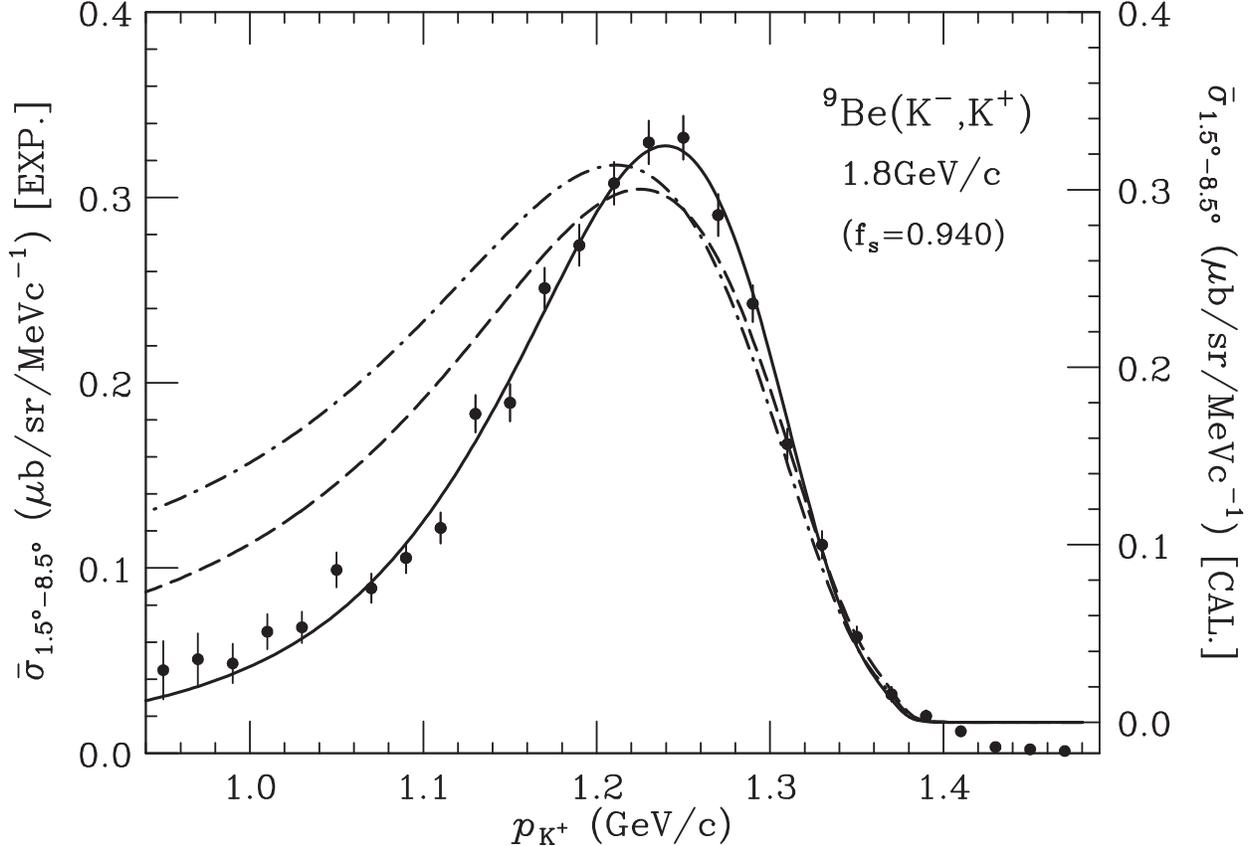}
\end{center}
\end{minipage}
\caption{\label{fig:5}
  Comparison of the calculated spectra 
  for $\bar{\sigma}_{1.5^\circ\mbox{-}8.5^\circ}$
  with the data of the $^9$Be($K^-$,~$K^+$) reaction 
  at $p_{K^-}=$ 1.80 GeV/$c$~\cite{Tamagawa00}, 
  using the WS potential with $V_0^\Xi =$ $-17$ MeV and $W_0^\Xi=$ $-$5 MeV. 
  Solid and dashed curves denote the spectra obtained by the optimal and 
  standard Fermi-averaged $K^-p \to K^+\Xi^-$ amplitudes for 
  $\overline{f}_{K^-p \to K^+\Xi^-}$, respectively. 
  A dot-dashed curve denotes the spectrum obtained by 
  $\beta (d \sigma/d \Omega)^{\rm av}_{\rm lab}=$ constant.   
  The spectra are folded with a detector resolution of 14.7 MeV FWHM. 
}
\end{figure}
%%%%%%%%%%%%%%%%%%%%%%%%%%%%%%%%%%%%%%%%%%%%%%%%%%%%%%%%%%%%%%%%%%

In a previous paper \cite{Harada20}, we emphasized the importance 
of the energy dependence of the $K^-p\to K^+\Xi^-$ amplitude of 
${\overline{f}}_{K^-p \to K^+\Xi^-}$ arising from the optimal Fermi-averaging 
procedure~\cite{Harada04} in the nuclear ($K^-$,~$K^+$) reaction. 
We discuss the calculated $\Xi^-$ QF spectra involving the energy dependence 
of ${\overline{f}}_{K^-p \to K^+\Xi^-}$ in comparison with the data of 
the $^9$Be($K^-$,~$K^+$) reaction in the BNL-E906 experiment.
To see the importance of the energy dependence of ${\overline{f}}_{K^-p \to K^+\Xi^-}$, 
we also estimate the spectrum in the DWIA 
using the ``standard'' Fermi-averaged cross section $(d\sigma/d\Omega)^{\rm av}_{\rm lab}$
for the $K^-p \to K^+\Xi^-$ reaction, 
which may be given by
\begin{eqnarray}
\Bigl( {d \sigma \over d \Omega} \Bigr)^{\rm av}_{\rm lab}
&=& \int d{\bm p}_N \rho(p_N) \Bigl( {d \sigma \over d \Omega} \Bigr)^{\rm free}_{\rm lab},
\label{eqn:10}
\end{eqnarray}
where $\rho(p_N)$ is a proton momentum distribution in the target nucleus, 
and  $(d \sigma/d \Omega)^{\rm free}_{\rm lab}$ is the differential cross section
for the $K^-p \to K^+\Xi^-$ reaction in free space.
This spectrum is proportional to $\beta (d \sigma/d \Omega)^{\rm av}_{\rm lab}$ 
indicating the energy dependence of $\beta$ 
whereas the value of $(d \sigma/d \Omega)^{\rm av}_{\rm lab}$ at each $\theta_{\rm lab}$ 
becomes constant in Eq.~(\ref{eqn:10}).
In Table~\ref{tab:table3}, 
we show the calculated values of $(d\sigma/d\Omega)^{\rm av}_{\rm lab}$ and 
$(d\sigma/d\Omega)^{\rm free}_{\rm lab}$ \cite{Harada20}. 
Figure~\ref{fig:5} displays the calculated $\Xi^-$ QF spectra 
obtained by the optimal and standard Fermi-averaged $K^-p \to K^+\Xi^-$ amplitudes 
in the $^{9}$Be($K^-$,~$K^+$) reaction at $p_{K^-}$= 1.8 GeV/$c$, together with 
the spectrum obtained by $\beta (d \sigma/d \Omega)^{\rm av}_{\rm lab}=$ constant, 
omitting the energy dependence of $\beta$. 
We find that the energy dependence of ${\overline{f}}_{K^-p \to K^+\Xi^-}$
acts on the shape and magnitude of the QF spectrum remarkably, 
and it makes its width narrower.
If we use a constant value for ${\overline{f}}_{K^-p \to K^+\Xi^-}$ in our calculations, 
the shape and magnitude of the calculated $\Xi^-$ QF spectrum cannot explain the data qualitatively. 
We show clearly 
that the optimal Fermi averaging for the $K^-p \to K^+ \Xi^-$ reaction provides 
a good description of the energy dependence of the $\Xi^-$ QF spectrum in 
the nuclear ($K^-$,~$K^+$) reaction \cite{Harada20}.
Therefore, we recognize that the optimal Fermi-averaged amplitudes for 
$\overline{f}_{K^-p \to K^+\Xi^-}$ is essential to explain the shape and magnitude 
of the spectrum including the $\Xi^-$ QF region with a wide energy range.
Thus it is required to extract information concerning the $\Xi$-nucleus 
potential carefully from the data of the experimental spectrum.

\section{Summary and Conclusion}
\label{summary}

We have studied phenomenologically the $\Xi^-$ production spectrum of 
the $^9$Be($K^-$,~$K^+$) reaction at 1.8 GeV/$c$ within the DWIA 
using the optimal Fermi-averaged $K^-  p \to K^+ \Xi^-$ amplitude. 
We have attempted to clarify properties of the $\Xi$-nucleus potential for $\Xi^-$-$^8$Li, 
comparing the calculated spectrum with the data of the BNL-E906 experiment. 
We have performed the $\chi^2$-fitting 
to the $N=$ 17 data points for $\bar{\sigma}_{1.5^\circ\mbox{-}8.5^\circ}$, 
varying the strength parameters of $V_0^\Xi$ and $W_0^\Xi$ in the WS potential. 

In conclusion, we show the weak attraction in the $\Xi$-nucleus potential 
for $\Xi^-$-$^8$Li,  
which provides the ability to explain the data for the $\Xi^-$ QF region 
in the $^9$Be($K^-$,~$K^+$) reaction at 1.8 GeV/$c$, 
consistent with analyses for previous experiments \cite{Khaustov00,Nakazawa15}.
The attraction of $V_0^\Xi =$ $-17 \pm 6$ MeV is favored within the WS potential, 
accompanied by the reasonable absorption of $W_0^\Xi =$ $-5$ MeV for
the $\Xi^-p \to \Xi^0n$, $\Lambda\Lambda$ transitions in nuclear medium, 
although it is difficult to determine the value of $W_0^\Xi$ from the data 
due to the insufficient resolution of 14.7 MeV FWHM.
The importance of the energy dependence of the Fermi-averaged $K^-p \to K^+ \Xi^-$ amplitude 
is confirmed by this analysis. 
The detailed analysis is also required for the J-PARC E05 experiment 
of the $^{12}$C($K^-$,~$K^+$) reaction at 1.8 GeV/$c$ \cite{Nagae18}.
This investigation is a subject for future research. 
\\

\begin{acknowledgments}
The authors thank Prof.~T.~Fukuda, Prof.~T.~Nagae, Prof.~Y.~Akaishi, 
Prof.~S.~Shinmura, and Dr.~A.~Dot\'e for many valuable discussions and comments. 
This work was supported by Grants-in-Aid for Scientific Research (KAKENHI)
from the Japan Society for the Promotion of Science
(Grant No.~JP20K03954).
\end{acknowledgments}

%-------------------------------------------------------------------


\clearpage

%%%%%%%%%%%%%%%%%%%%%%%%%%%%%%%%%%%%%%%%%%%%%%%%%%%%%%%%%%%%%%%%

\end{document}